\documentclass{article}

\usepackage[preprint]{neurips_2025}

\usepackage[utf8]{inputenc} 
\usepackage[T1]{fontenc}    
\usepackage{hyperref}       
\usepackage{url}            
\usepackage{booktabs}       
\usepackage{amsfonts}       
\usepackage{nicefrac}       
\usepackage{microtype}      
\usepackage{xcolor}         

\usepackage{graphicx}
\usepackage{subcaption}
\usepackage{amsmath}

\title{Textual forma mentis networks bridge language structure, emotional content and psychopathology levels in adolescents}

\author{%
    Alexis Carrillo\thanks{These authors contributed equally to this work.} \\
    CogNosco Lab\\
    Dept. of Psych. \& Cog. Sci. \\
    University of Trento, Italy \\
    \texttt{ae.carrilloramirez@unitn.it} \\
    \And
    Simon Friedrich Roske\footnotemark[1] \\
    CogNosco Lab\\
    Dept. of Psych. \& Cog. Sci. \\
    University of Trento, Italy \\
    \texttt{simon.roske@unitn.it} \\
    \And
    Rebeca Ianov-Vitanov \\
    MRC Cognition and Brain Sciences Unit \\
    University of Cambridge, UK \\
    \texttt{rebeca.vit@mrc-cbu.cam.ac.uk} \\
    \And
    Enrico Perinelli \\
    Dept. of Psych. \& Cog. Sci. \\
    University of Trento, Italy \\
    \texttt{enrico.perinelli@unitn.it} \\
    \And
    Alessandro Grecucci \\
    Dept. of Psych. \& Cog. Sci. \\
    University of Trento, Italy \\
    \texttt{alessandro.grecucci@unitn.it} \\
    \And
    Massimo Stella\thanks{To whom correspondence should be addressed: massimo.stella-1@unitn.it} \\
    CogNosco Lab\\
    Dept. of Psych. \& Cog. Sci. \\
    University of Trento, Italy \\
    \texttt{massimo.stella-1@unitn.it} \\
}

\begin{document}

\maketitle

\begin{abstract}
    We introduce a network-based AI framework for predicting dimensions of psychopathology in adolescents using natural language. We focused on data capturing psychometric scores of social maladjustment, internalising behaviors, and neurodevelopmental risk, assessed in 232 adolescents from the Healthy Brain Network. This dataset included structured interviews in which adolescents discussed a common emotion-inducing topic. To model conceptual associations within these interviews, we applied textual forma mentis networks (TFMNs)—a cognitive/AI approach integrating syntactic, semantic, and emotional word-word associations in language. From TFMNs, we extracted network features (semantic/syntactic structure) and emotional profiles to serve as predictors of latent psychopathology factor scores. Using Random Forest and XGBoost regression models, we found significant associations between language-derived features and clinical scores: social maladjustment (r = 0.37, p < .01), specific internalising behaviors (r = 0.33, p < .05), and neurodevelopmental risk (r = 0.34, p < .05). Explainable AI analysis using SHAP values revealed that higher modularity and a pronounced core–periphery network structure—reflecting clustered conceptual organisation in language—predicted increased social maladjustment. Internalising scores were positively associated with higher betweenness centrality and stronger expressions of disgust, suggesting a linguistic signature of rumination. In contrast, neurodevelopmental risk was inversely related to local efficiency in syntactic/semantic networks, indicating disrupted conceptual integration. These findings demonstrated the potential of cognitive network approaches to capture meaningful links between psychopathology and language use in adolescents.
\end{abstract}


\section*{Introduction}

The cognitive counterpart of language in the mind is a complex system in which ideas can be linked to convey meaning \cite{vitevitch2019network}. Whereas language exists in its own physical terms (written or digital traces), the cognitive processes that power language can only be investigated through theoretical models and tested indirectly through laboratory experiments \cite{aitchison2012words,harnad2017cognize}. A powerful cognitive model is the mental lexicon \cite{aitchison2012words,doczi2019overview}, an idealised cognitive system apt at acquiring, storing and processing knowledge expressed via language. The mental lexicon is modelled as a highly structured and dynamical system, storing, updating and accessing cognitive representations of concepts over time \cite{hills2022mind,vitevitch2019network,harnad2017cognize}. Over the last fifty years, a large body of empirical research has shown that the mental lexicon operates in associative ways\cite{doczi2019overview,aitchison2012words}: Concepts in the mental lexicon can be accessed via a variety of associations—semantic, syntactic, phonological, orthographic, emotional, and more. The structure of these associations influences how concepts are accessed and used \cite{doczi2019overview,harnad2017cognize}. For instance, the number of associations a given concept is involved into can influence its processing, memorisation and retrieval from memory \cite{vitevitch2022can}. 

The interplay between networked associations and language processing positions complex networks as a naturally powerful tool for investigating the mental lexicon. The blooming field of cognitive network science \cite{stella2022cognitive,Siew_2019_CogNetSci} bridges the associative structure of knowledge in the mental lexicon with cognitive and psychological phenomena related to language, as well as with all knowledge expressible through language. Applications of cognitive network science include the use of network topology to explain phenomena such as lexical processing \cite{vitevitch2022can}, creativity \cite{Kenett_2018_flexibility}, individual perceptions \cite{aeschbach2025mapping}, and social media perceptions \cite{abramski2024voices,stella2020text}, among others. 

A cognitive network can be defined as a complex network modelling associative knowledge in representational terms \cite{stella2024cognitive,Siew_2019_CogNetSci,castro2020contributions}. In other words, cognitive networks can encode the layouts of several types of similarities between concepts in the mental lexicon, including semantic, syntactic, phonological, orthographic, emotional and several others \cite{Semeraro_2025_EmoAtlas}. Multilayer cognitive networks can even encode multiple types of similarities at once among a single set of concepts \cite{stella2024cognitive}. These models are representational in nature—they capture structure but do not, on their own, perform computations, unlike artificial neural networks (ANNs) \cite{stella2020text}. Whereas in ANNs, nodes represent computational units used to approximate functions or solve regression and classification tasks, cognitive networks explicitly map the relationships between nodes that represent individual concepts—or, more simply, words.

Beyond computations, can the representations encoded in cognitive networks also reflect psychological constructs such as psychopathology levels (as defined in prior research \citep{Holmes_2021_threefactor})? This research question warrants careful consideration in connection with the so-called Deep Lexical Hypothesis \cite{cutler2023deep}. This hypothesis posits that mental states and psychological constructs can percolate through cognition to the extent that they alter an individual's language. In recent years, a rapidly growing body of evidence has supported this hypothesis (see Literature Review), particularly within the field of mental health. Several studies—despite not employing a network science approach—have identified distinct linguistic markers associated with varying levels of psychopathology.

\subsection*{Non-network linguistic markers of psychopathology}

Linguistic analysis reveals consistent associations between language patterns and psychopathology dimensions across various mental health conditions \cite{tausczik2010psychological}. 

Research shows that people experiencing depression tend to use language that is more self-focused and negatively valenced. For example, they often use a disproportionately high number of first-person singular pronouns (e.g., “I,” “me”) alongside more negative emotion words (e.g., “sad,” “hopeless”) in their writing \cite{Edwards2017}. The frequency of these linguistic markers has also been used in clinical samples to develop successful machine learning pipelines for detecting levels of depression from single-word expressions \cite{stade2023depression}. More specifically, both clinical interviews \cite{stade2023depression} and online posts on mental health forums \cite{al2018absolute} have shown that higher levels of depression are associated with increased use of first-person singular pronouns (e.g., “I”) and negative-emotion terms, suggesting that these verbal markers reliably reflect an individual’s depression state. In online forums, language from individuals experiencing depression was also found to contain a higher frequency of absolutist terms (e.g., “never,” “always,” “ever”).

Anxiety disorders, such as generalized anxiety disorder, were found to manifest linguistically through elevated usage of threat-related words (e.g., "danger," "risk") and future-oriented language \cite{Geronimi2015}. These findings align with evidence that anxious individuals show hyper-associative thinking around threats: In cognitive experiments, people high in anxiety more readily interpret ambiguous words in a threatening way, essentially “locking on” to threat-related meanings \cite{Richards1992}.

Always in the context of anxiety disorders, post-traumatic stress disorder (PTSD) is often accompanied by distinctive language features in trauma narratives. Studies of survivors’ written or spoken accounts find an overuse of visceral, sensory details (e.g., “scream,” “blood”) and disjointed or fragmented syntax when recounting a traumatic event \citep{Crespo2016}. This linguistic profile — rich in concrete imagery yet low in coherence — mirrors the intrusive memories and disorganised recollections characteristic of PTSD.  More recent work \citep{pugach2025uncovering}, identified a high rate of fluctuations of negative thoughs in PTSD patients, which is also related to PTSD severity (the higher the severity, the higher the fluctuations of negative thoughts).

In the context of eating disorders like anorexia nervosa, language use often centers obsessively on body image and food-related terms. Analyses of online forums and support groups for eating disorders have revealed an excessive focus on body shape/size and weight (words like “thin,” “weight,” “fat”) as well as on restrictive eating behaviors (e.g., “fast,” “starve,” “calories”). In one content analysis of thousands of tweets referencing eating disorder behaviors, 65$\%$ expressed a preoccupation with body shape or appearance\citep{Lyons2019}. This narrow vocabulary reflects the cognitive fixation seen in anorexia; indeed, the discourse of those affected is filled with recurrent references to dieting and body scrutiny. Network analyses of such language find that these terms form tightly interconnected clusters, indicating an almost rigid thought structure around body and food.

Bipolar disorder is marked by swings between manic and depressive states, and these shifts are echoed in language usage. During manic episodes, individuals’ speech tends to be rapid, excited, and filled with high-arousal positive words and grandiose or creative expressions (e.g., “ecstatic,” “brilliant,” “unbelievable”). In depressive episodes, by contrast, the same individuals’ language shifts to low-arousal negative content (e.g., “exhausted,” “worthless”) and slowed or minimal speech. Psycholinguistic analysis confirms this oscillation. Patients in mania show pressured speech with loose, free-associative leaps between ideas, whereas in depression their speech becomes subdued and sparse \citep{Sekulic2018}. 

Interestingly, many of the language indicators of psychopathology discussed above appear to generalize across different languages and cultures \citep{DeChoudhury2013}. This suggests that the link between certain word choices and underlying mental health conditions is not merely an artifact of English or any single language. In this sense, the Deep Lexical Hypothesis \citep{cutler2023deep} may point to a universal—or at least broadly generalizable—psycholinguistic signature that connects expressions within the mental lexicon to manifestations of psychopathology dimensions.

\subsection*{Network-based linguistic markers of psychopathology dimensions}

Fewer approaches have moved beyond individual words to identify markers of psychopathology dimensions in the associative relationships between concepts. Fatima and colleagues \cite{fatima2021dasentimental} showed that concept centrality in a network of memory recalls can enable an artificial intelligence (AI) model to predict levels of psychopathology dimensions—such as depression, anxiety, and stress—from word sequences, performing comparably to human raters. Analogously, Abramski and colleagues \cite{abramski2024voices} showed that networks of syntactic associations extracted from online posts by sexual assault survivors could reveal varying levels of distress, based on the degree of psychological distancing between the narrator and the events described. 

Mota and colleagues \cite{mota2012speech} built speech graphs from transcripts produced by individuals with psychosis. These graphs were built as networks of co-occurrences between words in texts. The authors found that individuals with schizophrenia produced more fragmented/disconnected graphs compared to a healthy control group. Subsequent work found that the connectivity of speech graphs was also anti-correlated with the extent of negative symptoms in patients with chronic psychosis \cite{mota2017thought}. This approach was importantly extended by Morgan and colleagues \cite{morgan2021natural} to include also other Natural Language Processing (NLP) markers next to speech graph connectivity. This study demonstrated that NLP-based measures and graph connectivity provided complementary insights, effectively distinguishing between individuals experiencing a first psychotic episode and those at clinical high risk for psychosis. Olah and colleagues \cite{olah2024towards} used speech graphs to show that connectivity, along with network features such as degree and diameter, can contribute to distinguishing between varying levels of psychotic risk and delusional ideation.

\subsection*{Research gaps and proposed framework: Textual forma mentis networks and AI}

Current literature highlights two relevant elements: (i) multiple studies indicate a link between the mental lexicon, as accessed to express language, and psychopathology dimensions; (ii) most network models used so far in mental health studies focused on modelling associative knowledge in the mental lexicon merely as co-occurrences between words in sentences. Although useful in detecting linguistic cohesiveness, co-occurrences include also spurious non-syntactic relationships between words \cite{aitchison2012words}, while missing specifications between words mentioned in remote parts of sentences \cite{stella2020text}. For instance, in the sentence "today I feel so much very — well, you know — sick!", the subject "I" and its syntactic complement "sick" are separated by several intervening words. As a result, their connection would not be captured by simple co-occurrence methods, despite a clear syntactic link. This limitation has led cognitive network science to move beyond co-occurring words and instead focus on syntactically related words within sentences — shifting the approach from co-occurrence networks to textual forma mentis networks (TFMNs, \cite{stella2022cognitive}).  

Taken together, these points highlight a crucial research gap in the use of cognitive networks—models of the associative structure of the mental lexicon—for predicting clinically assessed or psychometrically measured dimensions of psychopathology. This study builds on this growing body of empirical evidence while advancing it further by asking: Can the structure of the mental lexicon itself, rather than mere lexical access or speech-derived co-occurrence patterns, be used to predict levels of psychopathology levels/mental distress in a lab-controlled dataset?

Here, we propose using the cognitive framework of textual forma mentis networks \cite{stella2020text,stella2022cognitive,Semeraro_2025_EmoAtlas} for generating a network-based proxy of the associative knowledge encoded in the mental lexicon and expressed through language. TFMNs are constructed by identifying syntactic relationships among words within individual sentences in a given text. While such relationships are not explicitly written in the text, the mental lexicon is capable of parsing syntactic and semantic content to reconstruct meaning. Syntactic relationships include structures such as subject–verb agreement, verb–object dependencies, and noun modification by adjectives. TFMNs are created through syntactic parsing: a pre-trained AI model identifies these syntactic links and represents them as connections (edges) between corresponding words (nodes). The proximity of syntactically linked words within sentence structure can also be considered, with edges formed between concepts falling below a specified maximum syntactic distance (see Methods). These edges can be further enriched by semantic relationships and word-level features such as valence and emotional content. Thanks to their integration of syntactic, semantic, and emotional information, TFMNs offer a rich representation of associative knowledge as conveyed through text \cite{stella2022cognitive,Semeraro_2022_Emotional}.

\subsection*{Research questions and manuscript outline}

Our primary research aim is to evaluate the capability of textual forma mentis networks \cite{Semeraro_2025_EmoAtlas} to extract linguistically and psychologically relevant features for mental health research. By combining TFMNs with explainable AI, we investigate whether a cognitive data science framework can reliably detect linguistic markers that predict psychometric dimensions of psychopathology, using a dataset of over 200 expert-curated interviews with British adolescents.

Specifically, we apply a network-based machine learning approach to predict psychometric levels of social maladjustment, neurodevelopmental risk, and specific internalising symptoms. For key definitions of these psychometric dimensions see Methods - Target Variable Definitions. Then, through explainable AI, we explore whether variations in distress levels can be linked to interpretable network features within the cognitive framework of the mental lexicon. For example, network degree may serve as a proxy for the semantic or syntactic richness of conceptual associations. \cite{stella2020text} and a tendency for people to ruminate over the same concept in their interviews. The presence of high-degree nodes may, therefore, be interpreted as a potential indicator of rumination—a core symptom of depressive disorders \cite{al2018absolute}. Is this pattern supported by an AI pipeline that predicts psychopathology levels based on network degree? This study aims to explore research questions of this nature.

We outline our methodological framework in the Methods section, detailing the dataset and the construction of Textual Forma Mentis Networks (TFMNs), followed by an explanation of the explainable AI techniques and feature interpretations presented in the Results. The Discussion offers insights for data scientists, psychologists, and clinicians interested in innovative tools for assessing adolescent mental health. In light of the growing mental health crisis among adolescents, this research underscores the potential of language-based assessments for early detection and intervention, with implications for both educational and clinical settings.

\section*{Methods}
This study adopted the following framework: (1) using EmoAtlas \cite{Semeraro_2025_EmoAtlas}, construction of TFMNs from interview transcripts obtained from the Healthy Brain Network dataset; (2) extraction of network features and emotional z-scores from interviews; (3) exploratory analysis to identify initial relationships between predictor and target variables, the latter extracted from the HBN dataset with a p-factor model \cite{caspi2014pfactor}; (4) model training and testing using RFR and GBM; and (5) SHAP feature importance analyses to quantify and interpret feature contributions.

\subsection*{Data Source}
Data were sourced from the Healthy Brain Network (HBN) initiative \citep{Alexander_2017_HBN}, a large-scale project investigating the development of psychopathology in youth. The HBN initiative collected multimodal data, including psychiatric, behavioral, cognitive, and lifestyle information, as well as multimodal brain imaging data, from participants aged 5 to 21 years. Participants were recruited from the New York area, with a focus on families concerned about their child's psychiatric symptoms.

\subsection*{Selection of Interview Transcripts}
For this study, we used interview transcripts from a subset of the Healthy Brain Network (HBN) dataset. These transcripts were generated following a standardised protocol in which adolescents were interviewed after watching a short video designed to elicit emotional responses. The interviews were conducted by trained personnel from the Child Mind Institute. From the initial dataset of 664 participants in the first data release, 232 individuals with usable interview transcripts—suitable for linguistic analysis and TFMN construction—were selected. Participant age and biological sex were also included as predictor variables. While the sample was evenly split across biological sexes, participants ranged in age from early adolescence (13 years) to young adulthood (21 years).

\subsection*{Target Variable Definitions}
The target variables for the prediction models were three continuous, standardized factor scores representing latent dimensions of psychological distress. These scores were derived from a p-factor analysis with hierarchical structural equation modelling \cite{caspi2014pfactor} performed on a set of psychometric questionnaires. These were carefully selected by the authors to capture symptoms related to emotional regulation, social difficulties, and psychological distress—features commonly observed across a wide range of clinically diagnosed disorders in adolescence. The selection was informed by insights from previous research \cite{Holmes_2021_threefactor}. 

Our approach to detecting psychometric dimensions of mental distress is transdiagnostic: Distress manifests across a broad spectrum of symptoms — including persistent anxiety, depressive states, emotional dysregulation, and maladaptive coping strategies - which all can significantly impact academic performance and personal development \citep{laceulle2015structure}. Research consistently links such distress to a general psychopathology factor, or \textit{p}-factor, which represents an underlying higher-order dimension that accounts for the shared variance among various mental health disorders \citep{lahey2012there, caspi2014pfactor}. With this transdiagnostic approach, we included in this study the following questionnaires: the Revised Child Anxiety and Depression Scale - Parent Version (RCADS) \cite{de2002revised}, the Conners-3 Parent Short Form (CPSF) \cite{conners1998revised} and the Strengths and Difficulties Questionnaire (SDQ) \cite{goodman2009strengths}. These questionnaires are de facto standards for mental health assessments in youth \cite{Holmes_2021_threefactor} and can be used to account for symptoms of depression (RCADS), generalized anxiety disorder (RCADS), panic disorder (RCADS), social phobia (RCADS), separation anxiety (RCADS), obsessive-compulsive disorder (RCADS), aggression (CPSF), hyperactivity/impulsivity (CPSF), inattention (CPSF), maladaptive peer relations (CPSF), conduct problems (SDQ) and misregulated prosocial behavior (SDQ).

In a collection of questionnaires, factor analysis involves identifying clusters of responses that explain unique sources of variance within the dataset \cite{caspi2014pfactor}. Higher factor scores are associated with increased life-long impairment and compromised brain function, indicating more severe and pervasive psychological struggles \citep{Holmes_2021_threefactor,lahey2012there}. Through comparisons of factor solutions against randomised data, a confirmatory factor analysis of the data at hand resulted in a 3-factor solution, where we identified the following factors in the data:

\begin{itemize}
    \item \textbf{Social maladjustment:} Difficulties in social relationships and self-regulation, with higher scores indicating greater difficulties in blending or emotionally processing social events and bonds \citep{Holmes_2021_threefactor}.
    \item \textbf{Specific internalising or internalising behaviours:} Tendency to internalize emotional distress, derived from questionnaire items related to anxiety, depression, and panic. Higher scores reflect a greater tendency towards specific internalising behaviors \citep{Holmes_2021_threefactor,lahey2012there}.
    \item \textbf{Neurodevelopmental risk:} Impaired neurodevelopment, learning difficulties, and inattention. Higher scores indicate greater difficulties relative to impaired executive function and development \citep{caspi2014pfactor}.
\end{itemize}

\subsection*{Transforming texts in textual forma mentis networks}
\begin{figure}
    \centering
    \includegraphics[width=0.85\linewidth]{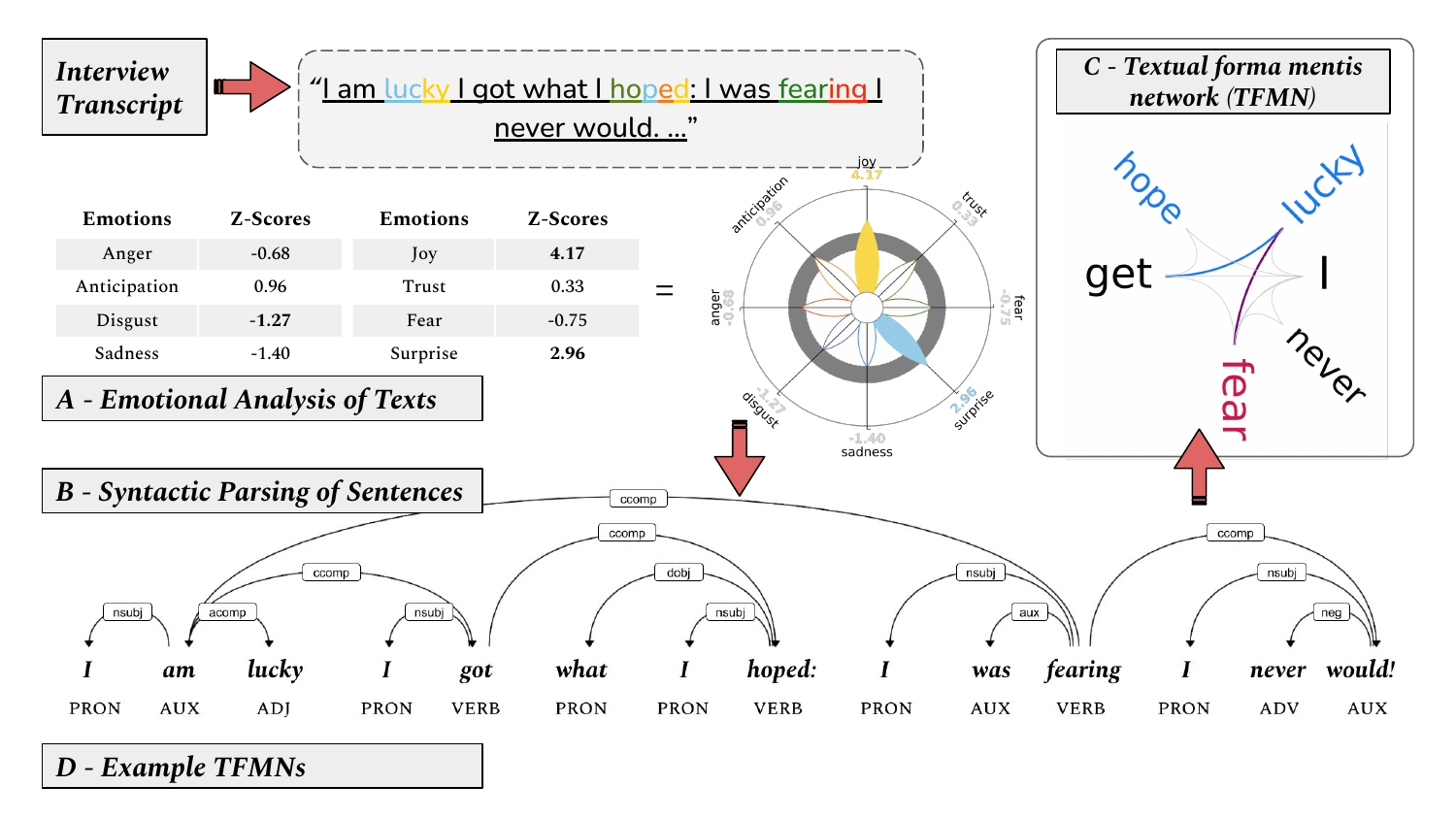}
    \includegraphics[width=0.425\linewidth]{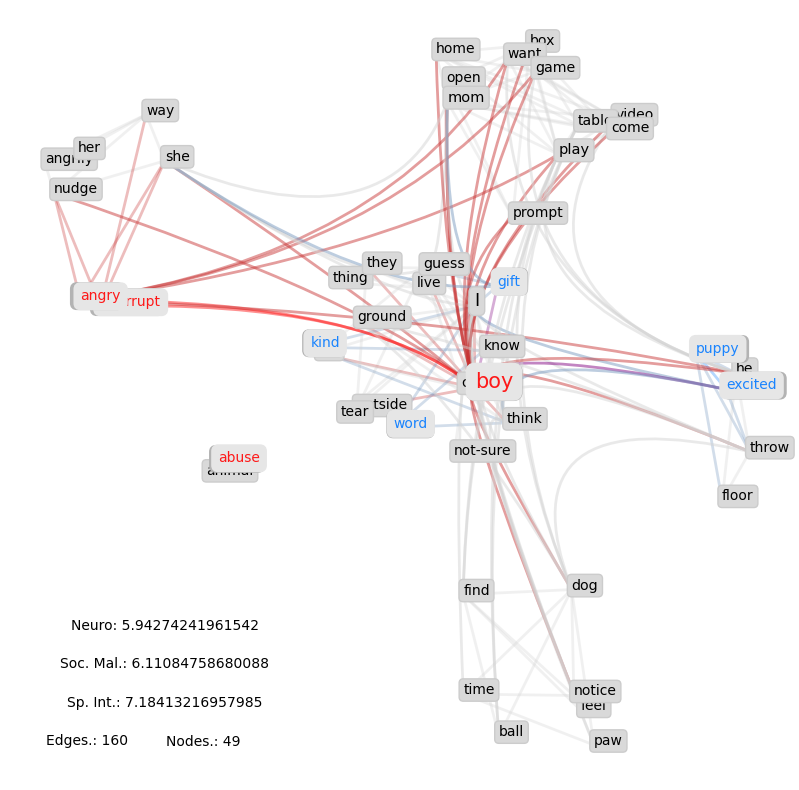}
    \includegraphics[width=0.425\linewidth]{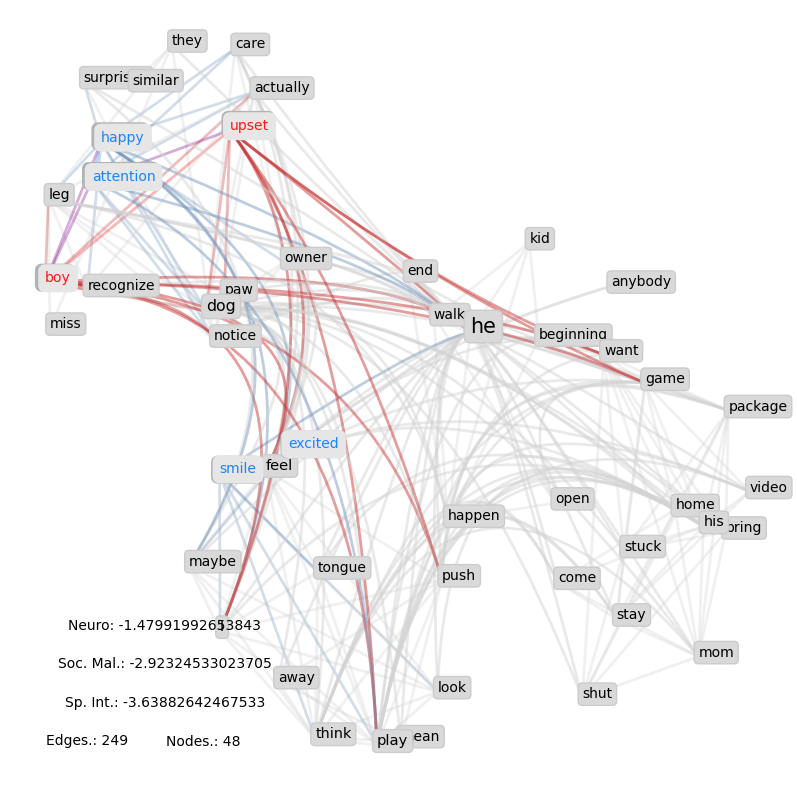}
    \caption{A,B,C: Algorithm steps for creating a textual forma mentis from an interview transcript. D: Two sample TFMNs illustrating differences in density, modular organization (more pronounced in b), and central node connectivity. Each TFMN represents the transcript text of a single interview provided by an individual in the dataset. In TFMNs, negative (positive) words are highlighted in red (cyan). Links indicate syntactic relationships and are cyan (red, purple) if between positive (negative, negative and positive) words.}
    \label{fig:combined_figures} 
\end{figure}

TFMNs were constructed using the EmoAtlas library in Python \cite{Semeraro_2025_EmoAtlas}. As introduced in previous work \cite{stella2020text,stella2022cognitive}, textual forma mentis networks are used to capture structural relationships within interview transcripts through a three-step process: syntactic parsing, integration of semantic relationships, and emotional profiling. Examples of textual forma mentis networks are reported in the illustrative Figure \ref{fig:combined_figures}.

First, syntactic parsing analyzed the grammatical structure of each sentence. An artificial intelligence model \cite{Semeraro_2025_EmoAtlas}, pre-trained on a large English corpus, enabled EmoAtlas to extract syntactic relationships and generate a syntactic tree for each sentence. To establish connections within the TFMNs, EmoAtlas measured the syntactic distance between non-stopwords in such syntactic tree: an edge was created between two non-stop words if their syntactic distance was less than or equal to a predefined parameter $k$, set to 4 in this study to focus on local syntactic associations within sentences. The selection of the syntactic distance parameter $k=4$ was informed by the cumulative distribution function of short-distance relations observed in the interview transcripts. A high proportion of syntactic relationships occurred at short distances: 68\% for $k=2$, approximately 86\% for $k=3$, and around 94\% for $k=4$. Consequently, $k=4$ was employed in the network construction, capturing a large majority of potential syntactic relationships within the sentences.

Second, semantic relationships, specifically synonyms, were integrated into the network using WordNet \cite{Miller_1995_WordNet}. WordNet is a lexical database providing a structured resource of semantic relations between words. The integration of these semantic links aimed to capture overlaps in meaning between words already connected within the syntactic network.

Third, emotional profiling was conducted using EmoLex \cite{mohammad2013crowdsourcing}, integrated within EmoAtlas. EmoLex identified words within the interview transcripts that evoke specific emotions. This analysis was performed at the word level, examining the emotions associated with the direct neighbors of a particular word within the TFMNs, and at the text level, assessing the overall presence and prominence of emotion-evoking words throughout the network constructed from an interview.

\subsection*{Extraction of network features from transcripts}
Graph-theoretic and emotion-based features were extracted from each TFMN for use as predictors of factor scores. These measures were calculated using functions from the Python library NetworkX. The following graph-theoretic features \citep{newman_2018_networks, Latora_2001_Efficient} were considered:

\begin{itemize}
    \item \textbf{Number of Nodes:} The total count of unique lexical concepts represented in the network.
    \item \textbf{Number of Nodes in the Largest Component:} The size of the largest connected subgraph within the overall network.
    \item \textbf{Maximum and Mean Degree:} The highest degree $k_{max}$ and the mean degree $\langle k \rangle$ among all nodes in the network. The degree $k_i$ of a node $i$ in a network represents the number of edges connected to that node. In TFMNs, this is the semantic/syntactic richness of a concept \cite{stella2022cognitive}. 
    \item \textbf{Core:} The largest subgraph in which every node has a degree of at least $k$. 
    \item \textbf{Mean Local Clustering Coefficient:} the average of the local clustering coefficients over all nodes $i$ in the network:
            \[ \bar{c} = \frac{1}{n} \sum_{i=1}^{n} c_i \]
    where $n$ is the total number of nodes in the network.
    The local clustering coefficient $c_i$ of a node $i$ quantifies the extent to which its neighbours are also connected to each other. This measure is the fraction of connections between neighbours of node $i$, whose count is $|e_i|$, and the total number of possible edges between neighbours. For a node with degree $k_i \ge 2$, $c_i$ is defined:
            \[ c_i = \frac{2 |e_i|}{k_i(k_i - 1)}\]
    If a node has a degree $k_i = 1$, the local clustering coefficient is typically defined as 0.
    \item \textbf{Mean Shortest Path:} the average smallest number of links connecting all pairs of nodes in the largest connected component of a network. The shortest path between two nodes $i$ and $j$ in a network is defined as the path with the minimum number of links connecting them.  
    \item \textbf{Diameter:} The maximum shortest path length between any pair of nodes in the network. 
    \item \textbf{Mean Closeness Centrality:} The mean closeness centrality $\bar{C}$ of the entire network is the average of the closeness centrality scores of all nodes $i$ in the network:
            \[ \bar{C} = \frac{1}{n} \sum_{i=1}^{n} C(i) \]
        where $n$ is the total number of nodes in the network.
        Closeness centrality $C(i)$ of a node $i$ quantifies its inverse average shortest path distance to all other $M$ reachable nodes in the network:
            \[ C(i) = \frac{M}{\sum_{j \neq i} d_{ij}} \]
        where $d_{ij}$ is the shortest path length between nodes $i$ and $j$. For disconnected graphs, this is often normalised by the number of reachable nodes.
    
    \item \textbf{Maximum Betweenness Centrality:} The highest betweenness centrality score among all nodes in the network. The betweenness centrality of a node $i$ is a measure of the number of times a node lies on the shortest path between other pairs of nodes in the network. The betweenness centrality $b_i$ of a node $i$ is given by:
    \[ b_i = \sum_{jk} \frac{n_{jk}(i)}{g_{jk}} \]
    where $g_{jk}$ is the total number of shortest paths from node $j$ to node $k$, and $n_{jk}(i)$ is the number of those shortest paths that pass through node $i$. The sum is over all pairs of distinct nodes $j$ and $k$ other than $i$.
        
    \item \textbf{Modularity:} $Q$ is a measure of the strength of division of a network into communities (also known as clusters or modules). It quantifies how well a particular division of the network into non-overlapping groups corresponds to the underlying network structure, with values typically ranging between -1 and 1, although in practice, values are usually non-negative. The modularity of a given partition of a network into $s$ communities is defined as:
        \[ Q = \frac{1}{2m} \sum_{i, j} \left[ A_{ij} - \frac{k_i k_j}{2m} \right] \delta(s_i, s_j) \]
        where:
        
        $A_{ij}$ is the adjacency matrix of the network (i.e., $A_{ij} = 1$ if there is an edge between nodes $i$ and $j$, and $A_{ij} = 0$ otherwise).\\
        $k_i$ is the degree of node $i$ (the number of edges connected to node $i$).\\
        $m$ is the total number of edges in the network.\\
        $s_i$ is the community to which node $i$ is assigned.\\
        $\delta(s_i, s_j)$ is the Kronecker delta function ($\delta(s_i, s_j) = 1$ if $s_i = s_j$, and $\delta(s_i, s_j) = 0$ otherwise).
    
    \item \textbf{Global Efficiency:} $E_{glob}(G)$ of a network $G$ is the average of the inverse shortest path lengths between all pairs of nodes in the network. Differently from closeness centrality, in efficiency one computes sums of inverse distances. 
    \item \textbf{Local Efficiency:}  $E_{loc}(i)$ of a node $i$ is defined as the efficiency of the subgraph consisting of the neighbours of $i$ with edges between them that exist in the original network. 

\end{itemize}

\subsection*{Emotional Profiling and Emotion Quantification}

Emotional profiling is the process of identifying and quantifying the emotional attributes associated with textual content, including individual words, concepts, or entire documents \cite{stella2022cognitive}. This typically involves the application of psychologically validated emotional lexicons \cite{mohammad2013crowdsourcing}. Within the context of TFMNs, emotional profiling is achieved by mapping these lexicon-derived emotional scores onto the nodes (words/concepts) of the network. This allows researchers to analyse the presence of specific emotions across the conceptual structure of the text, providing valuable insights into the affective tone and underlying sentiments associated with the expressed ideas. The integration of emotional profiling with the structural analysis of TFMNs enables a comprehensive understanding of how cognitive and emotional elements are intertwined in texts.

Figure \ref{fig:combined_figures} presents a visual representation of key TFMN characteristics and emotional profiles. Subfigures (a) and (b) illustrate two sample TFMNs, demonstrating variations in network density, modular organization, and central node connectivity, highlighting the structural diversity captured by this methodology. Subfigure (c) showcases an emotion flower, depicting the emotional content associated with the concept "dog" from a network text source. The size of each petal corresponds to the z-score of the emotion's intensity, with colored petals indicating statistically significant deviations ($|z| < 1.96  \text{,} \quad \alpha = 0.05$) from the mean, as delineated by the gray ring. This visualization effectively demonstrates how TFMNs can quantify and represent both structural and emotional information within textual data.

Emotional profiling was done using EmoAtlas \cite{Semeraro_2025_EmoAtlas} and detected the presence and intensity of eight primary emotions based on Plutchik's theory: anger, disgust, fear, trust, joy, sadness, surprise, and anticipation. $Z$-scores were used to compare the observed frequency of emotion words to a null model derived from EmoLex, allowing for the identification of significantly over- or underrepresented emotions.

Consider counting $n_e$ words as eliciting emotion $e$. We need to compare this count against a suitable null model. In EmoAtlas \cite{Semeraro_2025_EmoAtlas}, the null model is constructed through repeated random sampling from EmoLex \cite{mohammad2013crowdsourcing}. This preserves the language bias within the dataset (i.e. emotions do not have the same number of words encoding for them). The mean ($\langle r_e \rangle$) and standard deviation ($\sigma_{r_e}$) reference values for each emotion are obtained through the following process:

\begin{enumerate}
    \item  Identify $M$, the number of emotion-eliciting words in the target text. $M$ represents the count of words associated with at least one emotion in EmoLex.
    \item  Construct the null model by repeated sampling (uniformly at random). For a given set of words or text, $M$ words are randomly sampled from the emotion lexicon. This sampling is repeated $N$ times to create a distribution of expected emotion counts. The mean $\langle r_e \rangle$ for each emotion $e$ is calculated as the average of these counts across all $N$ samples, representing the expected frequency of each emotion in texts that are randomly assembled. On the same distribution, compute also the standard deviation $\sigma_{r_e}$ of counts.
\end{enumerate}

Once the mean $\langle r_e \rangle$ and standard deviation $\sigma_{r_e}$ are established for each emotion $e$, the $z$-score $Z_e$ is calculated as:

\[  Z_e = \frac{n_e - \langle r_e \rangle}{\sigma_{r_e}}\]

The $z$-score ($Z_e$) quantifies how far an observed count of emotion words ($n_e$) for a specific emotion deviates from the mean ($\langle r_e \rangle$) of the random null distribution. Once we have the $z$-score, fixing a significance level of $5\%$ lets the experimenter select emotional counts that are either substantially larger than random expectation ($Z_e > 1.96$, i.e. a text is rich in jargon eliciting that emotion) or substantially lower ($Z_e > 1.96$, i.e. a text is devoid of jargon eliciting that emotion). Higher $z$-scores also indicates stronger emotional intensities in transcripts, expressed by using more emotional words.

\subsection*{Feature standardisation and exploratory data analysis}

Prior to model training, all predictor features were standardized using a min-max scaler to ensure that features were on a common scale, facilitating model convergence and preventing features with larger ranges from dominating the learning process \cite{zhang2012ensemble}. To reduce dimensionality and mitigate potential multicollinearity, feature selection was performed. Specifically, features were selected from subsets of highly correlated features identified from a Pearson correlation matrix ($|r|>.1$).

Age and biological sex were included in all our models and participated within feature selection.

Exploratory data analysis was conducted to gain preliminary insights into the predictive potential of network features and emotional profiling. This initial exploration aimed to identify potential signals within the data and guide subsequent modeling efforts \cite{fatima2021dasentimental}. Pearson correlation coefficients ($r$) were computed to assess linear relationships between each network measure (e.g., modularity, core, centrality measures), each emotion score, and the three target psychological dimensions (Social Maladjustment, Specific internalising, and Neurodevelopmental risk). This provided an initial quantitative understanding of the strength and direction of linear associations between each linguistic feature and the different aspects of psychological distress. 

\subsection*{Machine Learning Models}
Two ensemble machine learning approaches \cite{zhang2012ensemble}, Random Forest Regression (RFR) and Gradient Boosting Machine (GBM), were employed to predict each dimension of mental distress/psychopathology levels from network features and emotional scores. These ensemble methods were selected for their effectiveness in capturing non-linearities in high-dimensional spaces, a characteristic particularly relevant to the multifaceted nature of psychological distress and the diverse feature space derived from TFMNs \cite{zhang2012ensemble}.

Hyperparameter tuning was conducted using grid search to identify optimal hyperparameter values for GBM and RFR models. The hyperparameter grids used are detailed below:

GBM Hyperparameter Grids:
\begin{itemize}
    \item Number of estimators: [5, 10, 25, 50, 100, 150]
    \item Learning rate: [0.1, 0.2, 0.3, 0.5, 0.7]
    \item Maximal depth: [2, 3, 5, 7, 9, 12, None]
    \item Maximum number of features: [log2, sqrt, None]
    \item Subsample: [0.5, 0.75, 1.0]
    \item Loss: [squared error, absolute error]
\end{itemize}

RFR Hyperparameter Grids:
\begin{itemize}
    \item Number of estimators: [5, 10, 25, 50, 100, 150]
    \item Maximal depth: [2, 3, 5, 7, 9, 12, None]
    \item Maximum number of features: [log2, sqrt, None]
    \item Maximum number of leaf nodes: [100, 150, None]
    \item Criterion: [squared error, friedman mse]
\end{itemize}

Gradient boosting and random forest models were trained using $k$-fold cross-validation ($k=4$) to predict scores on the three mental health dimensions: Social Maladjustment, Specific internalising, and Neurodevelopmental risk. Pearson correlation (largest absolute $r$ value, $p$ < .05) and Mean Absolute Error (MAE) between model output and actual values were used as performance measure. Models with higher significant correlations and lower MAE were considered to have better predictive performance and were therefore prioritized for further analysis and interpretation. Three sets of predictors were used: network features only, emotional profiling features only, and network and emotional features combined. To assess the extent to which model performance could be attributed to chance associations rather than meaningful relationships between the TFMN features and the psychological constructs, a permutation test was conducted \cite{zhang2012ensemble}. This involved randomly shuffling the values of target variables while preserving the original predictor variables.

\subsection*{Explainable AI and feature contribution analysis}
To analyse the overall impact and relationship of each predictor with the psychological dimensions, and to identify the most influential features, we used explainable AI \cite{xu2019explainable}. In Python, we computed Shapley scores indicating model performance based on variations of individidual predictor variables along a game-theoretic scenario (cf. \cite{xu2019explainable}). Beeswarm plots, mean SHAP value bar plots and heatmaps of SHAP values were created to provide a detailed view of how each feature affects predictions for individual psychological dimensions. This analysis aimed to enhance the interpretability of the models by relating network features and psychopathology levels. In fact, SHAP scores provided insights into the behavior of the fitted models by showing how variations in individual network features — from low to high values — influenced the predicted outcomes. For example, did higher modularity correspond to higher predicted values of maladaptive behaviour (Social Maladjustment)? Quantifying and exploring these feature–prediction relationships is a key contribution of the approach described in this manuscript.

\section*{Results}

Results are organised in this way: (1) exploratory results, (2) feature selection and model training, and (3) model performance and explainable AI patterns. 

\subsection*{Network features}
The exploratory analysis reveals significant correlations (at a $5\%$ significance level) between the considered factor scores and both network- and emotion features. Table \ref{tab:correlations_relevant} presents the Pearson correlation coefficients ($r$) derived from raw feature values. Mean betweenness centrality shows a positive correlation with Social Maladjustment ($r = 0.27$) and Neurodevelopmental Risk ($r = 0.15$), while exhibiting a negative correlation with Specific Internalising ($r = -0.28$). Mean degree demonstrates negative correlations across all three dimensions. Emotion-related features, such as Fear ($r = 0.11$), Anger ($r = 0.15$), Disgust ($r = 0.14$), and Surprise ($r = -0.13$), are correlated with Social Maladjustment. Surprise also shows a negative correlation with Specific Internalising ($r = -0.17$). 

Features including Joy, Trust, Sadness, Age, Sex, Positive Words, Diameter, and Reciprocity (RC) were excluded from the table due to correlations $|r| \leq 0.10$ across all three psychological dimensions.

The above findings convey preliminary evidence that both structural network features and emotional scores for transcripts can display some relationships with specific dimensions of mental distress. This is a key initial step prior to continuing our analysis with machine learning models - accounting for correlations and more complex patterns between predictors (network features and emotions) and target (factor scores) variables.

\begin{table}[htbp]
    \centering
    \caption{Pearson Correlations ($r$) Between TFMN Features and psychological dimensions.}
    \label{tab:correlations_relevant}
    \begin{tabular}{lcccc}
        \toprule
        & \textbf{Sp. Int.} & \textbf{Neuro.} & \textbf{Soc. Mal.} \\
        \midrule
        Fear & & & 0.11 \\
        Anger & & & 0.15 \\
        Surprise & -0.17 & & -0.13 \\
        Disgust & & & 0.14 \\
        Anticipation & -0.11 & & \\
        Max Degree & -0.19 & -0.12 & -0.17 \\
        Mean Degree & -0.27 & -0.22 & -0.27 \\
        Max Closeness & & -0.16 & \\
        Mean Deg. Assort. & -0.12 & 0.15 & 0.14 \\
        Mean Closeness & & & -0.11 \\
        | Components | & -0.13 & & \\
        Largest Comp. Ratio & -0.17 & & -0.13 \\
        Max Betweenness & -0.13 & & 0.19 \\
        Mean Betweenness & -0.28 & 0.15 & 0.27 \\
        | Largest Component | & -0.13 & & -0.11 \\
        | Nodes | & -0.11 & & -0.11 \\
        | Edges | & -0.17 & & -0.16 \\
        Density & & -0.15 & \\
        Mean Shortest Path & & 0.14 & 0.14 \\
        Max Clique & -0.18 & -0.15 & -0.20 \\
        Local Efficiency & -0.25 & -0.24 & -0.21 \\
        Global Efficiency & -0.20 & -0.18 & -0.19 \\
        Modularity & -0.14 & 0.15 & 0.18 \\
        | Core | & -0.17 & & -0.17 \\
        Core & -0.27 & -0.25 & -0.26 \\
        \bottomrule
    \end{tabular}\\
    \footnotesize{\textit{Note.} Correlations computed on raw data, $k=4$, $|r| \leq 0.10$ omitted. Sp. Int. = Specific Internalising, Neuro. = Neurodevelopmental Risk, Soc. Mal. = Social Maladjustment.}
\end{table}

\subsection*{Machine learning models predicting distress from network features and emotions}

Machine learning models were leveraged on the entire dataset of 232 transcripts/networks/vectors of psychometric features. A machine learning model (either GBM or RFR) was trained on the data with a 4-fold cross-validation, initially using all features as predictor variables. Features were ranked according to their total SHAP scores, capturing their influence over model performance. Starting from features with lower SHAP scores/feature importance, all values for a given feature were randomly reshuffled to measure a feature's influence over model performance during training/validation. If reshuffling feature values did not deteriorate model performance, the feature was discarded. This led to a hierarchical top-down feature selection process, whose outcomes are presented in Table \ref{tab:d4_unaltered}, which presents Pearson correlation coefficients ($r$), $p$-values, MAE, and the features selected by RFR and GBM models in 4-fold cross-validation rounds, predicting psychological dimensions.

For the Specific Internalising factor, the trained models of both random forests and XGBoost demonstrated significant positive correlations ($p < 0.05$), with GBM achieving a slightly higher correlation ($r = 0.33$) compared to RFR ($r = 0.31$). Similarly, both models showed significant positive correlations for the Neurodevelopmental Risk dimension ($p < 0.05$), with GBM again exhibiting a higher correlation ($r = 0.34$) than RFR ($r = 0.28$). In predicting the Social Maladjustment factor, GBM outperformed RFR, achieving a higher correlation ($r = 0.37$, $p < 0.01$) than RFR ($r = 0.32$, $p < 0.05$).
MAE was comparable between the two models across all three dimensions, with slight variations and considerably lower values for predicting Social Maladjustment in both RFR and GBM. The features selected by each model varied, indicating different feature importance assessments. GBM generally selected a wider range of features, particularly for Social Maladjustment. These results suggest that both models can effectively learn to predict the considered psychological dimensions starting from network and emotional features, with GBM demonstrating slightly superior performance, particularly in terms of predicting the Social Maladjustment factor.

\begin{table}[htbp]
    \centering
    \caption{Predictive Performance of RFR and GBM Models in 4-Fold Cross-Validation.}
    \label{tab:d4_unaltered}
    \begin{tabular}{lcccc}
        \toprule
        \textbf{Model} & \textbf{Correlation ($r$)} & \textbf{$p$-value} & \textbf{MAE} & \textbf{Features} \\
        \midrule
        \multicolumn{5}{c}{\textbf{RFR}} \\
        \midrule
        Specific Internalising & 0.31 & 0.03* & 2.39 & 2, 5, 6, D, J \\
        Neurodevelopmental Risk & 0.28 & 0.04* & 2.15 & 1, 2, 3 \\
        Social Maladjustment & 0.32 & 0.02* & 1.36 & 1, 2, U \\
        \midrule
        \multicolumn{5}{c}{\textbf{GBM}} \\
        \midrule
        Specific Internalising & 0.33 & 0.02* & 2.40 & 5, 6, D \\
        Neurodevelopmental  Risk & 0.34 & 0.02* & 2.10 & 1, 2, 3 \\
        Social Maladjustment & 0.37 & 0.01* & 1.34 & 1, 2, 3, 4, 7, U, G \\
        \bottomrule
    \end{tabular} \\    
    \footnotesize{\textit{Note.} Outcome variables were Specific Internalising, Neurodevelopmental Risk, and Social Maladjustment. Features included network measures (1 = modularity, 2 = core, 3 = local efficiency, 4 = maximum closeness centrality, 5 = age, 6 = mean betweenness centrality, 7 = maximum betweenness centrality) and emotions (S = sadness, J = joy, D = disgust, F = fear, A = anticipation, U = surprise, T = trust, G = anger). * $p < 0.05$.}
\end{table}

Restricting models to use only semantic/syntactic network features slightly deteriorates model performance, as reported in Table \ref{tab:d4_network_features}. For Specific Internalising, RFR achieved a significant positive correlation ($r = 0.30$, $p < 0.05$), while GBM showed a near-significant positive correlation ($r = 0.29$, $p = 0.05$). Both models demonstrated significant positive correlations for the Neurodevelopmental Risk dimension ($p < 0.05$), with GBM exhibiting a higher correlation ($r = 0.34$) compared to RFR ($r = 0.28$). In predicting Social Maladjustment, both models achieved significant positive correlations ($p < 0.05$), with RFR ($r = 0.27$) and GBM ($r = 0.27$) showing equal performance. MAE was comparable between the two models across all three dimensions, with slight variations and considerably lower values for predicting Social Maladjustment in both RFR and GBM. The features selected by each model varied, with RFR generally selecting fewer features. These results indicate that both models effectively predicted the psychological dimensions using only network features, with GBM showing a slight advantage in predicting the Neurodevelopmental Risk dimension. However, the restriction to network features only led to worse performances compared to models considering combinations of network and emotional features, as evident from the case of Specific Internalising in the GBM model, which provided statistically weaker results.

\begin{table}[htbp]
    \centering
    \caption{Predictive Performance of RFR and GBM Models Using Only Network Features.}
    \label{tab:d4_network_features}
    \begin{tabular}{lcccc}
        \toprule
        \textbf{Model} & \textbf{Correlation ($r$)} & \textbf{$p$-value} & \textbf{MAE} & \textbf{Features} \\
        \midrule
        \multicolumn{5}{c}{\textbf{RFR}} \\
        \midrule
        Specific Internalising & 0.30 & 0.04* & 2.34 & 1, 2 \\
        Neurodev. Risk & 0.28 & 0.04* & 2.15 & 1, 2, 3 \\
        Social Maladjustment & 0.27 & 0.04* & 1.37 & 1, 2, 3, 4 \\
        \midrule
        \multicolumn{5}{c}{\textbf{GBM}} \\
        \midrule
        Specific Internalising & 0.29 & 0.05 & 2.36 & 1, 2, 3 \\
        Neurodev. Risk & 0.34 & 0.02* & 2.10 & 1, 2, 3 \\
        Social Maladjustment & 0.27 & 0.05 & 1.39 & 2, 4, 5, 7 \\
        \bottomrule
    \end{tabular} \\    
    \footnotesize{\textit{Note.} Outcome variables were Specific Internalising, Neurodevelopmental Risk, and Social Maladjustment. Features included network measures (1 = modularity, 2 = core, 3 = local efficiency, 4 = mean betweenness centrality, 5 = global efficiency, 6 = maximum betweenness centrality, 7 = maximum closeness centrality). * $p < 0.05$.}
\end{table}

The performance of models using only emotional features is summarised in Table \ref{tab:d4_emotions}. For Specific Internalising, both RFR and GBM achieved positive correlations ($r = 0.27$), although these correlations were not statistically significant ($p = 0.09$). For the Neurodevelopmental dimension, RFR showed a non-significant negative correlation ($r = -0.07$, $p = 0.50$), while GBM showed a non-significant positive correlation ($r = 0.09$, $p = 0.44$). In predicting Social Maladjustment, both models achieved positive correlations ($r = 0.27$) but with non statistically significant p-values at the selected significance level ($p = 0.06$ for RFR and $p = 0.08$ for GBM).
MAE was lower for Social Maladjustment. The features selected by each model varied slightly, with GBM selecting a broader range of emotion features for Social Maladjustment. These results suggest that emotional features alone have very limited predictive power for the examined psychological dimensions. 

A comparison of Tables \ref{tab:d4_unaltered}, \ref{tab:d4_network_features} and \ref{tab:d4_emotions} indicates that combinations of semantic/syntactic networks features and emotional z-scores provide the best sets of features when predicting psychopathology levels. We interpret this result as an indication that not only emotions but also language structure and content can all convey useful information relative to psychopathology levels. To further investigate whether this pattern is genuinely present in the data or it should rather be attributed to skewed distributions of our predictor variables, we proceed by training sets of models with all features reshuffled to randomised values.

\begin{table}[htbp]
    \centering
    \caption{Predictive Performance of RFR and GBM Models Using Only Emotion Features.}
    \label{tab:d4_emotions}
    \begin{tabular}{lcccc}
        \toprule
        \textbf{Model} & \textbf{Correlation ($r$)} & \textbf{$p$-value} & \textbf{MAE} & \textbf{Features} \\
        \midrule
        \multicolumn{5}{c}{\textbf{RFR}} \\
        \midrule
        Specific Internalising & 0.27 & 0.09 & 2.39 & F, U \\
        Neurodev. Risk & -0.07 & 0.50 & 2.28 & S, D, J \\
        Social Maladjustment & 0.27 & 0.06 & 1.39 & G \\
        \midrule
        \multicolumn{5}{c}{\textbf{GBM}} \\
        \midrule
        Specific Internalising & 0.27 & 0.09 & 2.40 & F, U \\
        Neurodev. Risk & 0.09 & 0.44 & 2.22 & S, J \\
        Social Maladjustment & 0.27 & 0.08 & 1.38 & F, U, G \\
        \bottomrule
    \end{tabular} \\    
    \footnotesize{\textit{Note.} Outcome variables were Specific Internalising, Neurodevelopmental Risk, and Social Maladjustment. Features included emotions (S = sadness, J = joy, D = disgust, F = fear, A = anticipation, U = surprise, T = trust, G = anger).}
\end{table}

Table \ref{tab:d4_random} presents the results of our permutation test, displaying the performance metrics (Pearson correlations ($r$), $p$-values, MAE, and selected features) for both RFR and GBM models when applied to randomised datasets. Both RFR and GBM models exhibited substantially lower predictive performance compared to the models trained on the original data (Table \ref{tab:d4_unaltered}). Across all three psychological dimensions, the resulting correlations were non-significant ($p > 0.05$). For Specific Internalising, RFR showed a non-significant positive correlation ($r = 0.17$, $p = 0.25$), while GBM showed a non-significant negative correlation ($r = -0.22$, $p = 0.10$). For Neurodevelopmental Risk, RFR showed a non-significant negative correlation ($r = -0.13$, $p = 0.26$), and GBM showed a non-significant positive correlation ($r = 0.16$, $p = 0.22$). Similarly, for Social Maladjustment, both models yielded non-significant positive correlations: RFR at $r = 0.14$ ($p = 0.31$) and GBM at $r = 0.07$ ($p = 0.24$). 
These results, with substantially lower and non-significant correlations compared to the original analysis, indicate that the predictive accuracy observed in Table \ref{tab:d4_unaltered} is unlikely due to chance or spurious associations. Instead, these findings support the presence of meaningful relationships between the TFMN features and the psychological dimensions under investigation.

\begin{table}[htbp]
    \centering
    \caption{Predictive Performance of RFR and GBM Models with Randomized Outcome Variables.}
    \label{tab:d4_random}
    \begin{tabular}{lcccc}
        \toprule
        \textbf{Model} & \textbf{Correlation ($r$)} & \textbf{$p$-value} & \textbf{MAE} & \textbf{Features} \\
        \midrule
        \multicolumn{5}{c}{\textbf{RFR}} \\
        \midrule
        Specific Internalising & 0.17 & 0.25 & 2.22 & 2, 5, 6, D, J \\
        Neurodevelopmental Risk & -0.13 & 0.26 & 2.55 & 1, 2, 3 \\
        Social Maladjustment & 0.14 & 0.31 & 1.51 & 1, 2, U \\
        \midrule
        \multicolumn{5}{c}{\textbf{GBM}} \\
        \midrule
        Specific internalising & -0.22 & 0.10 & 2.91 & 2, 5, 6, D \\
        Neurodevelopmental Risk & 0.16 & 0.22 & 2.45 & 1, 2, 3 \\
        Social Maladjustment & 0.07 & 0.24 & 1.77 & 1, 2, 3, 4, 5, 7, D, A \\
        \bottomrule
    \end{tabular}\\
    \footnotesize{\textit{Note.} This analysis serves as a baseline comparison to Table \ref{tab:d4_unaltered}, which used the original (non-randomized) outcome variables. Features: 1 = modularity, 2 = core, 3 = local efficiency, 4 = max closeness centrality, 5 = age, 6 = mean betweenness centrality, 7 = max betweenness centrality, S = sadness, J = joy, D = disgust, F = fear, A = anticipation, U = surprise, T = trust, G = anger.}
\end{table}

\subsection*{Model comparison and feature importance analyses}

GBM generally demonstrated slightly better predictive performance than RFR across three psychological dimensions. Specifically, GBM achieved higher correlations with the observed scores for Specific internalising, Neurodevelopmental Risk, and Social Maladjustment (Table \ref{tab:d4_unaltered}). This enhanced performance was particularly evident in predicting Social Maladjustment, where GBM exhibited both a higher correlation and a lower MAE compared to RFR. Furthermore, GBM tended to select a broader range of features. Based on these observations, GBM was selected as the primary model for subsequent analysis and interpretation in terms of SHAP scores.

To elucidate the predictive mechanisms of the GBM models, a SHAP analysis was performed on the GBM models trained using the scaled feature space ($[-5, 5]$), which included network measures derived from TFMNs and emotion-based profiling metrics. By examining the distribution and magnitude of SHAP values, we identified the most influential features driving the model's predictions and interpreted their impact on the prediction of each psychological dimension. Specifically, we assessed how variations in feature values affected the model's output, revealing the direction and strength of each feature's influence.

\subsubsection*{Explainable AI for Social Maladjustment} 

\begin{figure}[htbp]
    \centering
    \caption{SHAP analysis of feature contributions for predicting Social Maladjustment with GBM. The dataset was scaled to $[-5, 5]$. (a) Beeswarm plot showing the relationship between observed feature values (y-axis) and their impact on the model output (x-axis). (b) Mean absolute SHAP values for each feature, indicating their average importance across the dataset. (c) Heatmap of SHAP values for all instances, with blue indicating negative contributions and red indicating positive contributions to the model output.}
    \begin{subfigure}[b]{\textwidth}
        \centering
        \includegraphics[width=0.5\textwidth]{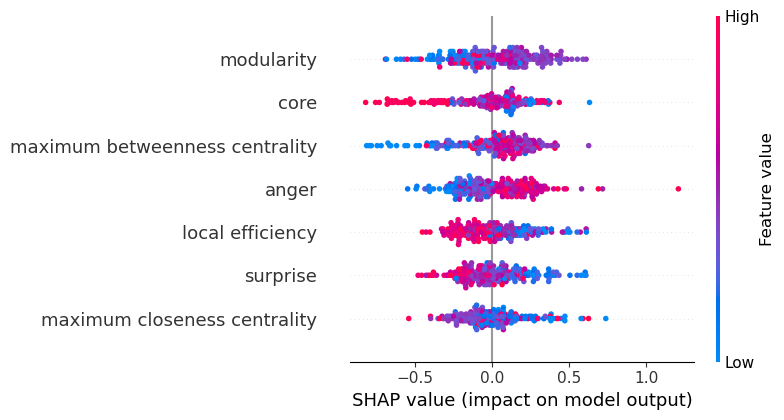}
        \caption{Beeswarm plot of feature values and SHAP values.}
        \label{fig:beeswarm_soc_mal_gbm}
    \end{subfigure}
    \vfill
    \begin{subfigure}[b]{0.49\textwidth}
        \centering
        \includegraphics[width=0.9\textwidth]{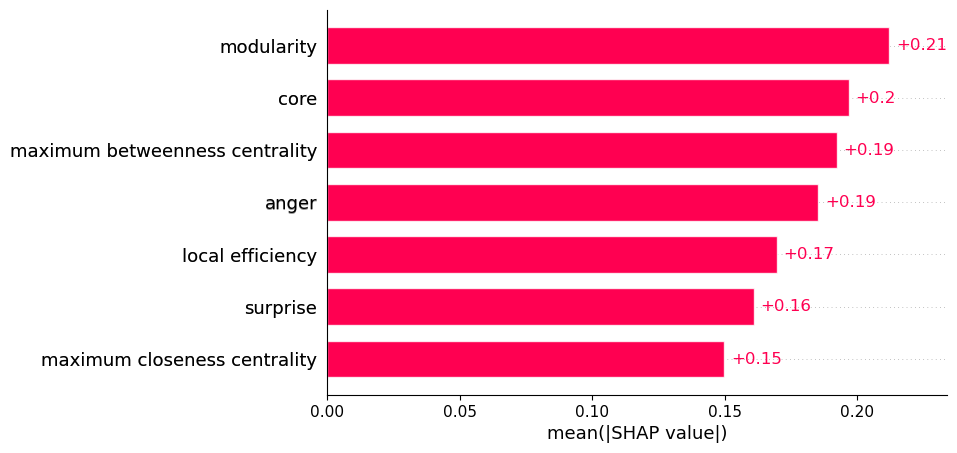}
        \caption{Mean absolute SHAP values.}
        \label{fig:barplot_soc_mal_gbm}
    \end{subfigure}
    \hfill
    \begin{subfigure}[b]{0.49\textwidth}
        \centering
        \includegraphics[width=0.9\textwidth]{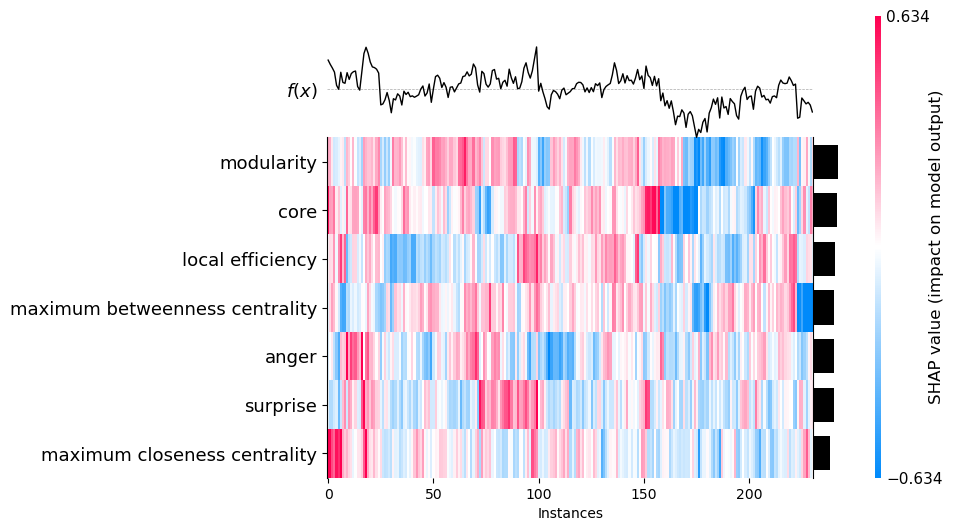}
        \caption{Heatmap of SHAP values.}
        \label{fig:heatmap_soc_mal_gbm}
    \end{subfigure}
    \label{fig:shap_discussion_soc_mal_gbm}
\end{figure}

The GBM model for Social Maladjustment identified the following features as relevant for prediction: core structure, maximum betweenness centrality, disgust, anger, modularity, local efficiency, maximum closeness centrality, and age. The SHAP trends for this model are reported in Figure \ref{fig:shap_discussion_soc_mal_gbm}.

Modalurity was the most influential network feature for predicting Social Maladjustment, cf. Fig. \ref{fig:shap_discussion_soc_mal_gbm} (b). As evident from the heatmap, Fig. \ref{fig:shap_discussion_soc_mal_gbm} (c), higher values of modularity corresponded to higher predicted value for Social Maladjustment. This means that the model is capturing a correspondence between syntactic/semantic networks being more compartmentalised when coming from individuals with higher values of Social Maladjustment. Since modularity indicates the presence of communities \cite{newman_2018_networks} and communities in TFMNs reflect topics of conversations \cite{Semeraro_2022_Emotional}, this pattern indicates a more prominent organisation in topics during interviews for individuals with higher levels of Social Maladjustment.

Core structure also demonstrated a clear feature-to-target relationship, with larger core structures associated with lower predicted Social Maladjustment. The remaining features contributed with similar average impacts, in descending order of importance. For other network features, lower values of maximum betweenness centrality were associated with increased predicted Social Maladjustment in a subset of individuals, indicating that reduced influence of certain nodes as bridges within the TFMN may contribute to higher levels of social maladjustment. 

Interestingly, increased expression of anger and decreased expression of surprise were both generally associated with higher predicted Social Maladjustment. While high values of mean betweenness centrality, maximum closeness centrality, and core occasionally showed positive associations with predicted Social Maladjustment, these relationships were not consistently observed across all individuals. Furthermore, lower values of local efficiency and core were strongly associated with higher predicted Social Maladjustment. This suggests that less efficient local information processing within the TFMN and a less densely interconnected core structure may indicate higher scores for the  Social Maladjustment dimension.

\subsubsection*{Explainable AI for Specific Internalising}

\begin{figure}[htbp]
    \centering
    \caption{SHAP analysis of feature contributions for predicting Specific Internalising with GBM. The dataset was scaled to $[-5, 5]$. (a) Beeswarm plot showing the relationship between observed feature values (y-axis) and their impact on the model output (x-axis). (b) Mean absolute SHAP values for each feature, indicating their average importance across the dataset. (c) Heatmap of SHAP values for all instances, with blue indicating negative contributions and red indicating positive contributions to the model output.}
    \begin{subfigure}[b]{\textwidth}
        \centering
        \includegraphics[width=0.5\textwidth]{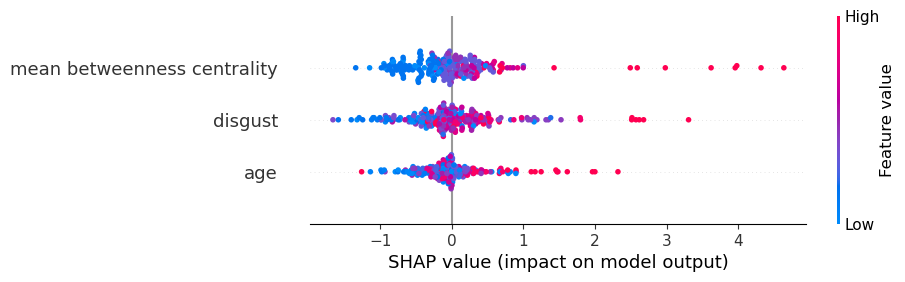}
        \caption{Beeswarm plot of feature values and SHAP values.}
        \label{fig:beeswarm_spint_gbm}
    \end{subfigure}
    \vfill
    \begin{subfigure}[b]{0.49\textwidth}
        \centering
        \includegraphics[width=0.9\textwidth]{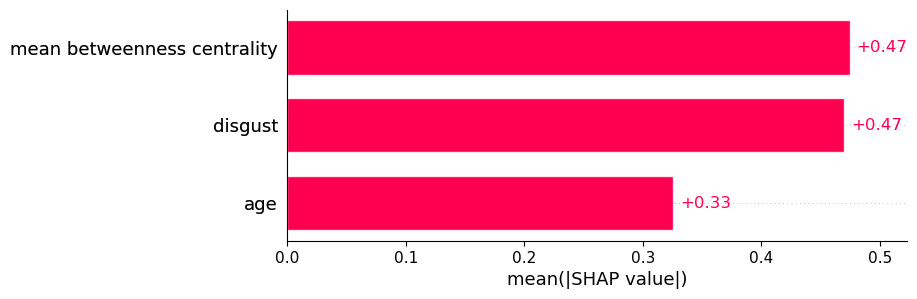}
        \caption{Mean absolute SHAP values.}
        \label{fig:barplot_spint_gbm}
    \end{subfigure}
    \hfill
    \begin{subfigure}[b]{0.49\textwidth}
        \centering
        \includegraphics[width=0.9\textwidth]{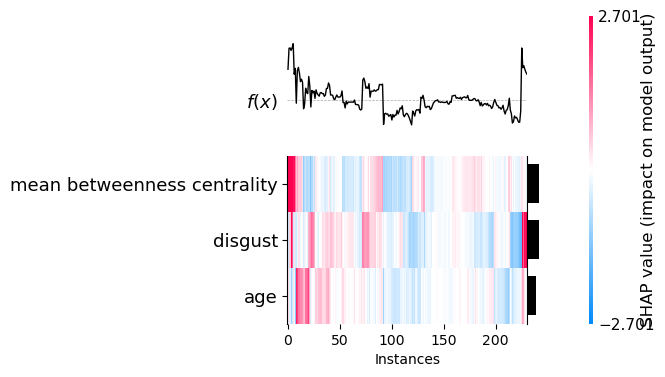}
        \caption{Heatmap of SHAP values.}
        \label{fig:heatmap_spint_gbm}
    \end{subfigure}
    \label{fig:shap_discussion_spint_gbm}
\end{figure}

The GBM model for Specific internalising (Figure \ref{fig:shap_discussion_spint_gbm}) identified mean betweenness centrality, disgust, and age as the most influential predictors. Although the overall SHAP scores identified several outliers, higher values of these features tended to increase the model's predicted Specific Internalising scores, with mean betweenness centrality having the strongest effect, followed by disgust and age.

Large SHAP values for these features were observed only in a minority of cases, indicating that their strong contribution to the prediction was limited to specific instances. For most individuals, the predictions were closer to the mean, suggesting a less pronounced influence of these features. Notably, the SHAP heatmap, cf. Fig. \ref{fig:heatmap_spint_gbm} (c), revealed a distinct two-cluster separation for disgust, primarily driving the distinction between positive and negative predictions. This suggests that disgust played a particularly crucial role in the later stages of the model's learning process, effectively differentiating between individuals with higher and lower scores on the Specific Internalising factor. 

These findings highlight the complex interplay of network structure, emotional expression, and age in predicting the Specific Internalising factor. While certain features may exert a strong influence in specific cases, the overall predictive pattern appears to be more nuanced and less reliant on consistently high values of individual features.

\subsubsection*{Explainable AI for Neurodevelopmental Risk}

\begin{figure}[htbp]
    \centering
    \caption{SHAP analysis of feature contributions for predicting Neurodevelopmental Risk with GBM. The dataset was scaled to $[-5, 5]$. (a) Beeswarm plot showing the relationship between observed feature values (y-axis) and their impact on the model output (x-axis). (b) Mean absolute SHAP values for each feature, indicating their average importance across the dataset. (c) Heatmap of SHAP values for all instances, with blue indicating negative contributions and red indicating positive contributions to the model output.}
    \begin{subfigure}[b]{\textwidth}
        \centering
        \includegraphics[width=0.5\textwidth]{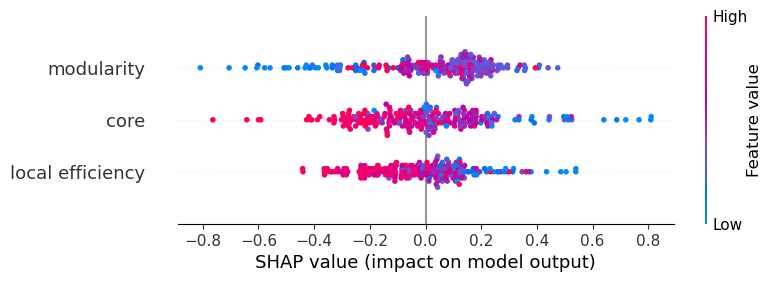}
        \caption{Beeswarm plot of feature values and SHAP values.}
        \label{fig:beeswarm_neuro_gbm}
    \end{subfigure}
    \vfill
    \begin{subfigure}[b]{0.49\textwidth}
        \centering
        \includegraphics[width=0.9\textwidth]{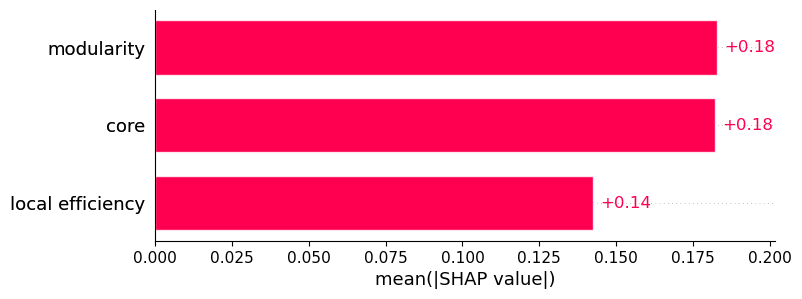}
        \caption{Mean absolute SHAP values.}
        \label{fig:barplot_neuro_gbm}
    \end{subfigure}
    \hfill
    \begin{subfigure}[b]{0.49\textwidth}
        \centering
        \includegraphics[width=0.9\textwidth]{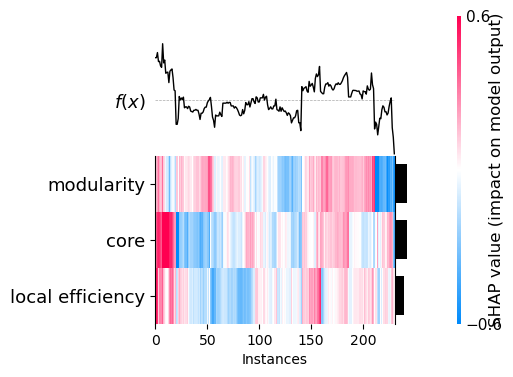}
        \caption{Heatmap of SHAP values.}
        \label{fig:heatmap_neuro_gbm}
    \end{subfigure}
    \label{fig:shap_discussion_neuro_gbm}
\end{figure}

The GBM model for Neurodevelopmental Risk (Figure \ref{fig:shap_discussion_neuro_gbm}) identified core structure, local efficiency, and modularity as the most influential predictors. Interestingly here, differently from the other two dimensions, age did not survive to the feature selection process. This distinction might indicate patterns that accumulate early on during development and persist across different ages. 

Core structure and local efficiency exhibited an inverse relationship with the predicted Neurodevelopmental Risk scores, where higher values of these features were associated with lower predicted scores. This suggests that a more densely interconnected core structure and greater local efficiency in information processing within the TFMN may indicate lower levels of neurodevelopmental challenges.

In contrast, the relationship between modularity and the predicted Neurodevelopmental scores was more complex. Lower modularity values were more frequently associated with negative SHAP values, indicating that lower modularity may contribute to higher predicted Neurodevelopmental scores. However, the impact of higher or neutral modularity values was less pronounced.

These findings suggest that the GBM model for Neurodevelopmental relies on a complex interplay of network features, with core structure and local efficiency acting as protective factors, while the influence of modularity appears to be more nuanced and context-dependent.

\section*{Discussion} 
This study shows that cognitive features expressed with language and reconstructed via AI can be used to predict psychometric levels of psychopathology in adolescents. This reconstruction is realised here by TFMNs \cite{stella2020text}, cognitive networks reconstructing semantic, syntactic and emotional aspects of associative knowledge recalled from the mental lexicon \cite{doczi2019overview}. Such knowledge was analysed here as transcripts of 232 interviews, from the Healthy Brain Network \cite{Alexander_2017_HBN}, where individual teenagers, guided by mental health experts, expressed themselves on emotional subjects. Through an innovative combination of network science, emotion analysis and interpretable AI, the analysis performed here provides insights into the syntactic, semantic, and emotional markers associated with social maladjustment, internalising difficulties and neurodevelopmental risk of adolescents.

Social maladjustment is an inability to conform to social norms, resulting in conduct problems and hampering peer relations and prosocial behavior \cite{Holmes_2021_threefactor,caspi2014pfactor}. When predicting psychometric levels of social maladjustment,  XGBoost models highlighted a substantial impact of several network measures relative to community structure and network connectedness. Medium-to-high modularity and core structure, but also higher betweenness and lower efficiency/closeness, all predicted higher social maladjustment values. These elements \cite{newman_2018_networks} all convey the presence of a marked organisation of concepts in tightly connected clusters - communities or core - that are integrated with each other (higher betweenness) while remaining distinct (giving rise to peripheral nodes with lower maximal closeness centrality). In general, TFMN clusters can reflect different topics of conversation \cite{Semeraro_2025_EmoAtlas} and the current findings, highlighted by machine learning here, might indicate for a tendency of individuals with higher social maladjustment to organise their conversations in highly thematic topics of conversation. Our analysis did not investigate the semantic topics of these conversations but rather their emotional content: Individuals with higher levels of social maladjustment were found by the model to express stronger levels of anger and weaker levels of surprise. Anger is an emotion expressing aggressiveness, with key neural correlates relative to expressing one's self \cite{sorella2021anger}, and its occurrence in teenagers with social maladjustment has been widely documented, cf. \cite{ariyazangane2022relation}. Anger can create feedback loops expressing maladjustment while also deteriorating social relationships when overexpressed \cite{ariyazangane2022relation}. Thus, our findings are consistent with past works and indicate a tendency for teenagers with social maladjustment to express anger in semantically/syntactic clustered topics during emotional interviews.

Specific internalising difficulties cover a broad spectrum of emotional dysregulation and learning issues \cite{Holmes_2021_threefactor,caspi2014pfactor}. Compared to social maladjustment, our explainable AI pipeline highlighted a smaller set of features: higher betweenness centrality, age, and stronger levels of disgust resulted in higher predicted psychometric values of the Specific Internalising factor. Betweenness captures how many bottlenecks are present in the layout of shortcuts in a given network \cite{newman_2018_networks}. In TFMNs bottlenecks are semantic/syntactic associations being used as shortcuts to link larger sets of concepts \cite{stella2022cognitive}. Always going through the same negatively-valenced associations (e.g. expressing disgust) for linking concepts might be an indication of rumination, a tendency for thought processes to fixate on the same associations, topics or ideas \cite{watkins2020reflecting}. Rumination is also a symptom of a wide variety of clinical conditions linked with emotional dysregulation \cite{stade2023depression,al2018absolute}. Alternatively, disgust, as an emotion of closure against/rejection of the external environment \cite{mohammad2013crowdsourcing}, might be also the expression of issues with internalising concepts and emotions. Interestingly, the AI model indicates that specific internalising difficulties are developmentally relevant, getting more problematic in older teenagers, which is in agreement with previous findings \cite{Holmes_2021_threefactor}. 

Neurodevelopmental factors are mostly relative to attention deficit disorders, conduct disorders, and learning difficulties \cite{Holmes_2021_threefactor,caspi2014pfactor}. Our AI framework identified medium-to-high modularity, medium-to-low coreness and lower local efficiency as being relative to higher predicted values of neurodevelopmental risk. Analogously to social maladjustment, these network features identified an organisation of concepts in well-integrated yet loosely defined topics/clusters, here devoid of emotional content. The structural organisation of language in topics might reflect speech cohesiveness, which was observed to be a predictor of psychosis levels in past works \cite{morgan2021natural,mota2017thought,mota2012speech}.

The present findings suggest that language-based features—operationalised through textual forma mentis networks \citep{Semeraro_2025_EmoAtlas} and grounded in cognitive network science—hold promise as accessible and scalable markers for capturing both general and specific dimensions of adolescent psychopathology. Our study builds upon and integrates prior work \cite{mota2012speech,pennebaker2003psychological,morgan2021natural,olah2024towards}, contributing further evidence in support of the Deep Lexical Hypothesis \cite{cutler2023deep}, which posits a systematic relationship between language structure and psychopathology.

We argue that the investigation of language and mental health should go beyond simple bag-of-words approaches, and this is where network science could play a pivotal role. Cognitive networks like TFMNs provide a systematic and theoretically grounded method \cite{stella2022cognitive} for extracting meaningful features from unstructured textual data. This approach goes beyond simple bag-of-words by incorporating syntactic dependencies, semantic associations, and emotional profiling directly into the feature engineering process, allowing data scientists to derive features inherently linked to linguistic structure and psychological concepts (e.g., linking betweenness and rumination together with psychometric data via explainable AI). A key advantage is the "glass box" nature of TFMNs, contrasting with traditional "black box" machine learning models \cite{rudin2019stop}. This transparency can enable data scientists and mental health experts to understand the specific linguistic patterns and conceptual associations that contribute to the prediction of psychological constructs. 

Our study is not devoid of limitations. The limited sample size of this work led to cross-validation being used on the whole dataset, thus reducing model generalisability. This choice was pursued not only because of the need to maximise training (i.e. the number of different and complex instances observed by the AI model) but also because of a general lack of interest in predictability of unseen data. The approach pursued in this study was not the creation of regression models of any use in clinical practice. Instead, our work aimed at using explainable AI to investigate patterns in the current data, linking network structure and emotional content as accurately as possible with psychometric dimensions. In presence of future, larger datasets from the Healthy Brain Network \cite{Alexander_2017_HBN} or affine initiatives, the framework implemented here could be easily expanded to feature nested cross-validation and generalisation tasks with the aim of building predictive AI models via future research. Another limitation of our approach is that we considered only the text of transcripts produced by interviewees. This approach neglected the content of the emotional video seen by all participants and the semantic content of questions posed to the latter. Although questions did not vary across participants, some of their speech conformed to the questions, e.g. using "it" rather than "dog" because that noun was used in the question already. Once future network models capable of handling co-referencing are developed, the same network features explored in this study could be readily re-applied and evaluated against the current findings. It is important to note that although TFMNs do not account for coreferencing, the features of modularity and coreness identified in this study are not driven by a few isolated nodes. Instead, they emerge from a global integration of multiple interconnected concepts. Therefore, the current findings are not contingent on coreferencing, though future analyses could incorporate it to enrich the results—particularly if researchers wish to explore how individuals refer to semantically or syntactically central entities in the video stimulus.

\subsection*{Conclusions and Future Research}

The Deep Lexical Hypothesis proposes that language reflects underlying cognitive and emotional states. Our findings support this idea, showing that Textual Forma Mentis Networks (TFMNs) can successfully predict psychometric levels of psychopathology in adolescents based on transcripts of emotionally charged interviews. By capturing both emotional and structural network features, TFMNs offer a unique lens through which to investigate the interplay between language and internal psychological states. Understanding the factors that modulate this relationship holds significant potential for advancing clinical and educational applications.

\begin{ack} 
The authors gratefully acknowledge Sarah Morgan for extensive discussions that contributed to the foundational ideas of this paper. MS also acknowledges the IDSAI Turing-Exeter Award for support in exploring a preliminary version of the dataset.
\end{ack}

\bibliographystyle{abbrv}
\bibliography{references}

\newpage

\appendix

\renewcommand{\thefigure}{S\arabic{figure}}
\setcounter{figure}{0}

\section{Supplementary Information}

This Section contains additional information for the random forest classifier and explainable AI.

\subsection{RFR SHAP Values for Social Maladjustment}

This Section contains SHAP analyses of feature contributions for predicting Social Maladjustment with RFR, as reported in Figure S1. Model output is explained as an additive contribution of features. SHAP values represent individual sample contributions and averaged feature importance. The input dataset was scaled to [-5, 5]. (a) Beeswarm plot illustrating the relationship between observed feature values (y-axis) and their impact on the RFR model's prediction (x-axis). (b) Bar plot displaying mean absolute SHAP values for each feature, indicating their average importance across the dataset. (c) Heatmap visualizing SHAP values for all instances, with color gradients representing feature contributions: blue indicates negative contributions (decreasing the model's prediction), and red indicates positive contributions (increasing the model's prediction). The RFR model's prediction is shown in the top header, and mean SHAP value contributions are represented by dark bars on the right side.

\begin{figure}[htbp]
    \centering
    \caption{SHAP analyses of feature contributions for predicting Social Maladjustment with RFR. Model output is explained as an additive contribution of features.}
    \begin{subfigure}[b]{\textwidth}
        \centering
        \includegraphics[width=0.5\textwidth]{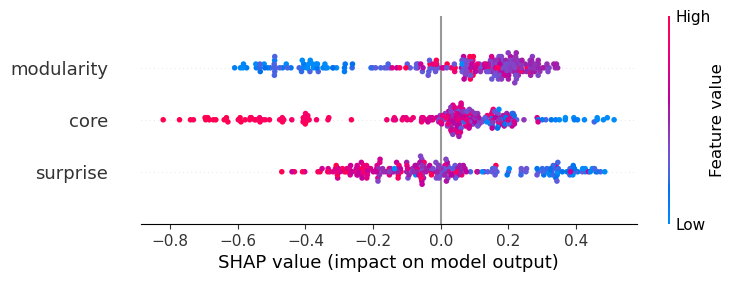}
        \caption{Beeswarm plot of feature values and SHAP values.}
        \label{fig:beeswarm_soc_mal_rfr}
    \end{subfigure}
    \vfill
    \begin{subfigure}[b]{0.49\textwidth}
        \centering
        \includegraphics[width=0.9\textwidth]{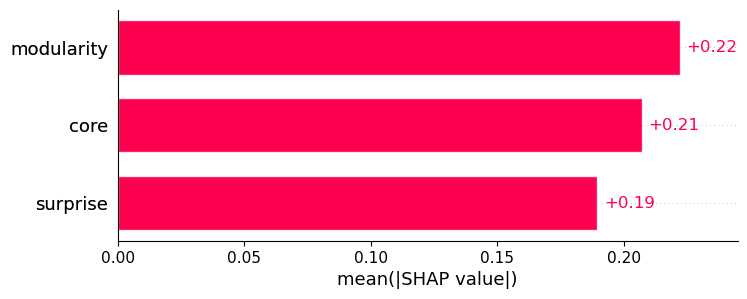}
        \caption{Mean absolute SHAP values.}
        \label{fig:barplot_soc_mal_rfr}
    \end{subfigure}
    \hfill
    \begin{subfigure}[b]{0.49\textwidth}
        \centering
        \includegraphics[width=0.9\textwidth]{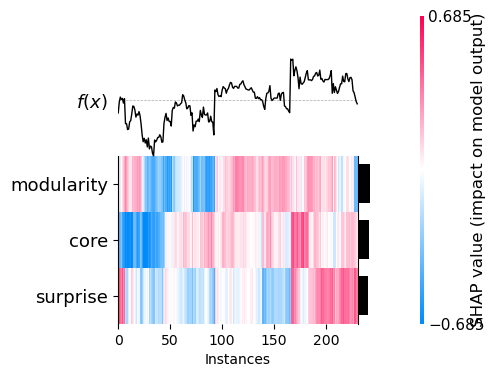}
        \caption{Heatmap of SHAP values.}
        \label{fig:heatmap_soc_mal_rfr}
    \end{subfigure}
    \label{fig:shap_discussion_soc_mal_rfr}
\end{figure}

\subsection{RFR SHAP Values for Specific Internalising}

This Section contains SHAP analyses of feature contributions for predicting Specific Internalising with RFR, as reported in Figure S2. Model output is explained as an additive contribution of features. SHAP values represent individual sample contributions and averaged feature importance. The input dataset was scaled to $[-5, 5]$. (a) Beeswarm plot illustrating the relationship between observed feature values (y-axis) and their impact on the RFR model's prediction of Specific Internalising (x-axis). (b) Bar plot displaying mean absolute SHAP values for each feature, indicating their average importance across the dataset. (c) Heatmap visualizing SHAP values for all instances, with color gradients representing feature contributions: blue indicates negative contributions (decreasing the model's prediction), and red indicates positive contributions (increasing the model's prediction). The RFR model's prediction is shown in the top header, and mean SHAP value contributions are represented by dark bars on the right side.

\begin{figure}[htbp]
    \centering
    \caption{SHAP analyses of feature contributions for predicting Specific Internalising with RFR. Model output is explained as an additive contribution of features.}
    \begin{subfigure}[b]{\textwidth}
        \centering
        \includegraphics[width=0.5\textwidth]{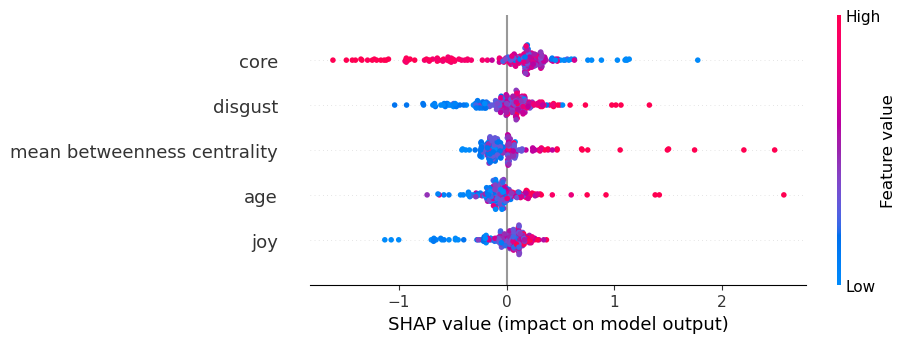}
        \caption{Beeswarm plot of feature values and SHAP values.}
        \label{fig:beeswarm_spint_rfr}
    \end{subfigure}
    \vfill
    \begin{subfigure}[b]{0.49\textwidth}
        \centering
        \includegraphics[width=0.9\textwidth]{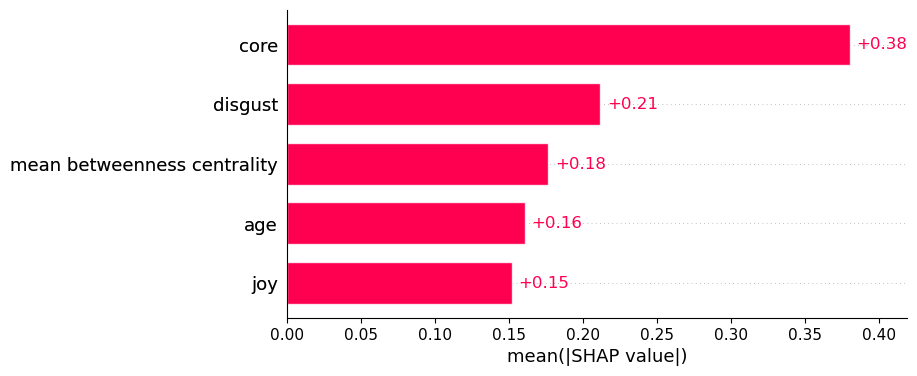}
        \caption{Mean absolute SHAP values.}
        \label{fig:barplot_spint_rfr}
    \end{subfigure}
    \hfill
    \begin{subfigure}[b]{0.49\textwidth}
        \centering
        \includegraphics[width=0.9\textwidth]{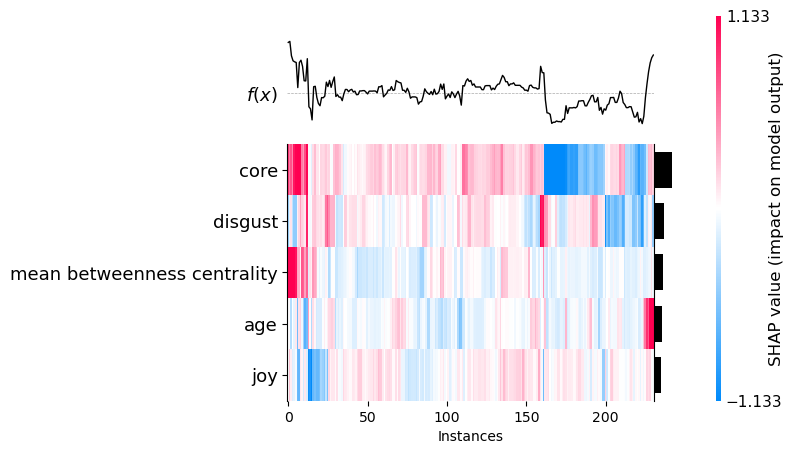}
        \caption{Heatmap of SHAP values.}
        \label{fig:heatmap_spint_rfr}
    \end{subfigure}
    \label{fig:shap_discussion_spint_rfr}
\end{figure}

\subsection{RFR SHAP Neurodevelopmental Risk}

This Section contains SHAP analyses of feature contributions for predicting Neurodevelopmental Risk with RFR, as reported in Supplementary Figure S3. Model output is explained as an additive contribution of features. SHAP values represent individual sample contributions and averaged feature importance. The input dataset was scaled to $[-5, 5]$. (a) Beeswarm plot illustrating the relationship between observed feature values (y-axis) and their impact on the RFR model's prediction of the Neurodevelopmental Risk factor (x-axis). (b) Bar plot displaying mean absolute SHAP values for each feature, indicating their average importance across the dataset. (c) Heatmap visualizing SHAP values for all instances, with color gradients representing feature contributions: blue indicates negative contributions (decreasing the model's prediction), and red indicates positive contributions (increasing the model's prediction). The RFR model's prediction is shown in the top header, and mean SHAP value contributions are represented by dark bars on the right side.

\begin{figure}[htbp]
    \centering
    \caption{SHAP analyses of feature contributions for predicting Neurodevelopmental Risk with RFR.}
    \begin{subfigure}[b]{\textwidth}
        \centering
        \includegraphics[width=0.5\textwidth]{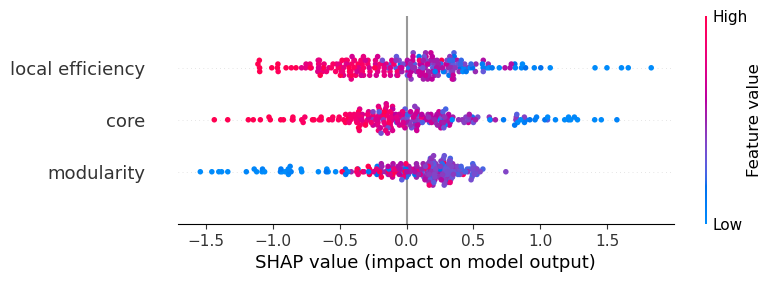}
        \caption{Beeswarm plot of feature values and SHAP values.}
        \label{fig:beeswarm_neuro_rfr}
    \end{subfigure}
    \vfill
    \begin{subfigure}[b]{0.49\textwidth}
        \centering
        \includegraphics[width=0.9\textwidth]{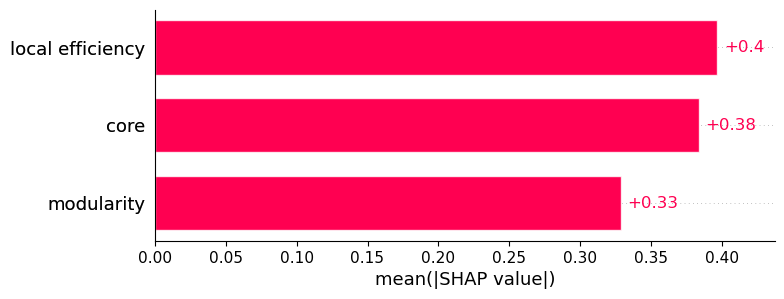}
        \caption{Mean absolute SHAP values.}
        \label{fig:barplot_neuro_rfr}
    \end{subfigure}
    \hfill
    \begin{subfigure}[b]{0.49\textwidth}
        \centering
        \includegraphics[width=0.9\textwidth]{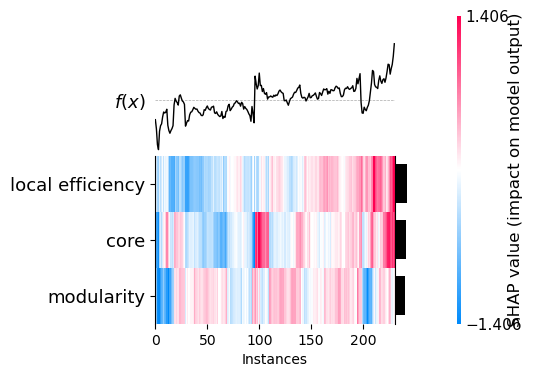}
        \caption{Heatmap of SHAP values.}
        \label{fig:heatmap_neuro_rfr}
    \end{subfigure}
    \label{fig:shap_discussion_neuro_rfr}
\end{figure}

\newpage
\section{Acronyms} 

\begin{itemize}
    \item Textual Forma Mentis Network (TFMN)
    \item Healthy Brain Network (HBN)
    \item Word-Emotion Association Lexicon (EmoLex)
    \item Gradient Boosting Machine (GBM)
    \item Random Forest Regression (RFR)
    \item SHapley Additive exPlanations (SHAP)
    \item Mean Absolute Error (MAE)
\end{itemize}

\end{document}